\documentstyle[twocolumn,prl,aps,epsfig,amsmath,floats,times]{revtex}

\begin{document}
\draft
\title{Super-lattice, rhombus, square, and hexagonal standing waves \\
in magnetically driven ferrofluid surface} 
\author{Hyun-Jae Pi, So-yeon Park, Jysoo Lee, and Kyoung J. Lee\cite{email1}}
\address{National Creative Research Initiative Center for
Neuro-dynamics \\ 
and Department of Physics, Korea University, Seoul 136-701, Korea}
\date{Dec 2, 1999} 
\maketitle

\begin{abstract}
Standing wave patterns that arise on the surface of ferrofluids by
(single frequency) parametric forcing with an ac magnetic field are
investigated experimentally.  Depending on the frequency and amplitude
of the forcing, the system exhibits various patterns including a
superlattice and subharmonic rhombuses as well as conventional
harmonic hexagons and subharmonic squares.  The superlattice arises in
a bicritical situation where harmonic and subharmonic modes collide.
The rhombic pattern arises due to the non-monotonic dispersion
relation of a ferrofluid.
\end{abstract}



The phenomena of self-organization in spatially extended
nonequilibrium systems have attracted much attention lately in the
scientific community world-wide \cite{ch93}.  
One of the recent scientific issues in this
regard is to investigate the system in which a finite number of
unstable (spatial and temporal) modes interact each other, while the
past studies have mostly focussed on the limiting cases of either
pattern forming systems where a single mode is excited or systems that
are highly turbulent with many modes.  Although the mode-coupling
interactions have been a long standing issue in various fields ranging
from material science to plasma physics, their significance in
nonequilibrium pattern formation is only begun to be understood.

In a recent experiment of parametrically driven
surface waves, Edwards and Fauve used a two-frequency excitation
scheme and were able to generate a quasi-crystalline standing wave
pattern with 12-fold symmetry \cite{Fauve93}.  In subsequent
experiments on a similar experimental setup, Kudrolli {\it et
al.}~observed two different types of superlattice pattern (``type-I''
and ``type-II'') in addition to the 12-fold symmetric
quasi-crystalline pattern \cite{Gollub98}.  More recently, Arbell and
Fineberg report yet another kinds of superlattice pattern (``SSS'' and
``2MS'') \cite{Fineberg98}.  All these unusual patterns are obtained
in two-frequency forced Faraday experiments in which two {\it
externally} excited modes interact (i.~e. near bicriticality).  Very
recently, however, Wagner {\it et al.}~have demonstrated that such
bicritical condition can be achieved even by a single frequency
forcing for a viscoelastic liquid \cite{Knorr99-1} or for a normal
Newtonian fluid in an extreme driving condition (very shallow filling
depth and large shaking elevations) \cite{Knorr99-2}.  New types of
superlattice pattern are revealed in those systems.  The existence of
these unusual patterns all together suggests a large class of
previously unanticipated spatio-temporal patterns with multiply
interacting modes.

In the present paper, we take a step in this direction by experimental
study on a vertically oscillating ferrofluid that is driven by a
sinusoidal magnetic field.  There are two reasons for choosing the new
system over previously investigated ones.  First of all, the driven
ferrofluid system can be easily brought near to a bicritical situation
with a single frequency forcing -- the harmonic mode as well as
subharmonic mode can be generated by an ac magnetic field with a small
driving frequency ($<30$ Hz) and a small magnetic field ($<100$ gauss)
\cite{Salin91}.  This contrasts with the case of
mechanically driven Newtonian fluids in which either a complex
two-frequency forcing scheme or an extreme driving condition is
necessary to achieve a bicritical situation \cite{mwwak97}.  The other
reason is that ferrofluid exposed to a magnetic field can exhibit a
non-monotonic dispersion relation \cite{r85}.  Due to the
non-monotonicity the neural stability curve for the excitation of
standing waves can have multiple minima \cite{Riecke97,Rehberg98-1},
thus, several spatial modes can be excited simultaneously.

\begin{figure}
\centerline{\psfig{figure=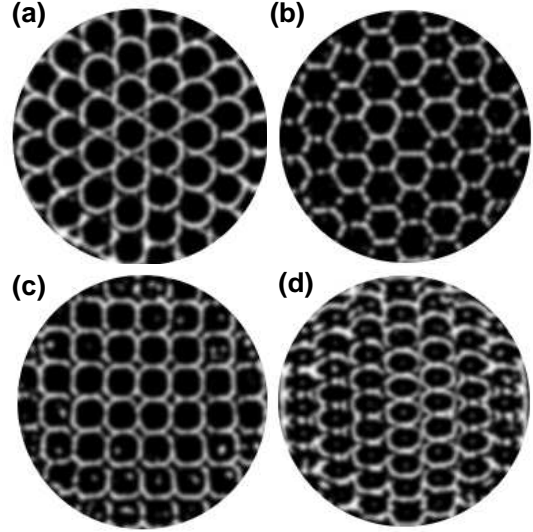,width=0.4\textwidth}}
\caption{Regular patterns observed on the surface of a driven
ferrofluid in a cylindrical container: (a) harmonic hexagon ($f = 6.0$
Hz), (b) subharmonic superlattice ($f = 6.7$ Hz), (c) subharmonic
square ($f = 11.0$ Hz), and (d) subharmonic rhombus ($f = 20.0$ Hz).
The total magnetic field is $H(t) = H_{0} + \Delta H \sin (2 \pi f
t)$, where $H_{0}/H_{c}$ and $\Delta H/H_{c}$ are fixed at $0.94$ and
$0.22$, respectively.  The diameter of the viewing area is $60$ mm.}
\label{fig1}
\end{figure}

As expected, the driven ferrofluid system is quite rich to reveal a
variety of standing wave patterns as shown in Fig.~\ref{fig1}.  Two
unusual patterns, (1) superlattice with a mixed-mode oscillation and
(2) subharmonic rhombic pattern, as well as usual subharmonic square
and harmonic hexagonal patterns are observed.  The observed
superlattice is similar but different from any of the previously
observed ones in its shape and temporal evolution.  The rhombic
pattern is the first example of its kind ever demonstrated in
experiments on nonequilibrium pattern formation.  

The ferrofluid is a colloidal suspension of magnetic powder stabilized
by screened electrostatic repulsion \cite{r85}.  We use a
commercially available ferrofluid (EMG901, Ferrofluidics) [density
$\rho = 1.53$ gm/ml, surface tension $\sigma = 29.5$ g/${\rm s}^{2}$,
initial magnetic susceptibility $\chi = 3.00$, magnetic saturation
$M_{s} = 600$ gauss, dynamic viscosity $\eta = 10$ cp, yielding a
critical field of the static Rosensweig-instability $H_c = 79.73$
gauss].  A cylindrical Teflon container containing the ferrofluid
(depth = $5$ mm, diameter = $85$ mm) is placed in the center of a pair
of Helmholtz-coils (Cenco instruments) with a inner (outer) diameter
of $100$ mm ($300$ mm).  The magnetic field is monitored by means of a
hall probe (F. W. Bell Inc., Model 4048), and the spatial variation of
the field strength from the center to the outer rim is within 3\%.  An
ac signal is generated from a home-built synthesizer-board and
amplified by a linear amplifier driving the total magnetic field of
$H(t)=H_{0}+{\Delta}H\sin(2{\pi}ft)$.  $H_{0}$ is the static field,
${\Delta}H$ is the amplitude of ac component, and $f$ is the driving
frequency.  $H_{0}$ is fixed at $0.94H_{c}$ for all cases, while
${\Delta}H$ and $f$ are used as a control parameter.

The fluid surface is illuminated by two arrays of concentric LED rings
(diameter $100$ mm and $120$ mm, respectively) located $230$ mm above
the surface.  The flat surfaces either above or below the level of the
surrounding fluid appear white, while the non-flat surfaces that
scatter the light away from the camera appear black.  The patterns are
imaged using a high speed $256 \times 256$ pixel CCD camera (Dalsa,
CA-D6-0256W) located $370$ mm above the surface with a fast frame
grabber (Matrox, Meteor-II).  Maximum frame acquisition rate is $955$
frame/s, enabling us to photograph distinct phases of moving pattern.

\begin{figure}
\centerline{\psfig{figure=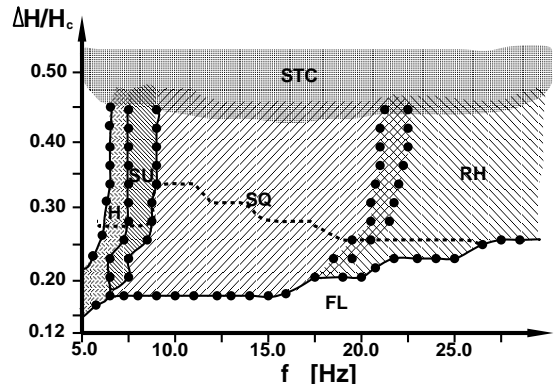,width=0.4\textwidth}}
\caption{Phase diagram of standing wave patterns revealed in a
magnetically driven ferrofluid with $H_{0} = 0.94 H_{c}$: flat surface
(FL), hexagon (H), superlattice (SU), square (SQ), and rhombus (RH).
The thin shaded area between SQ and RH is bistable (i.~e.~SQ and RH
coexist).  Spatio-temporal chaotic state (STC) is generally observed
at large $\Delta H$.  Crossing the dashed line toward the STC state,
the crystalline states start to acquire defects.}
\label{fig2}
\end{figure}

The phase diagram shown in Fig.~\ref{fig2} summarizes our results.
Upon increasing sequence of ${\Delta}H$, the flat surface can be
unstable to various crystalline states depending on the value of
driving frequency $f$.  Squares and rhombuses occur over a wide range
of parameter space, whereas hexagons and superlattice form only in a
narrow range at low driving frequency. Spatio-temporal chaotic state
(STC) gradually forms with increasing number of defects as ${\Delta}H$
increases.  Essentially, the same phase diagram is obtained with a
square container.  

\begin{figure}
\centerline{\psfig{figure=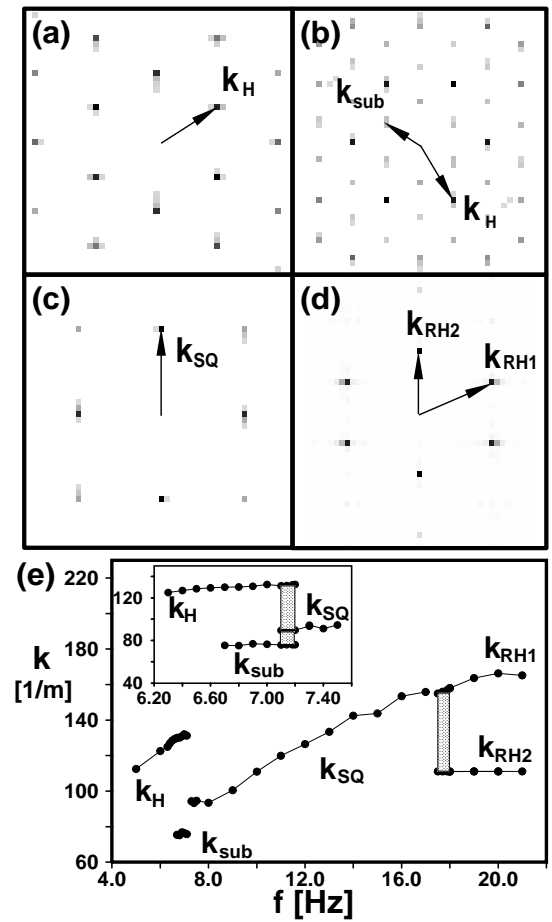,width=0.4\textwidth}}
\caption{Fourier transforms of the regular patterns shown in
Fig.~\ref{fig1}: (a) hexagon, (b) superlattice, (c) square, and (d)
rhombus.  The magnitude of wavevectors associated with different
patterns vs $f$ are shown in (e).  The insets shows the details of the
transition among hexagon, superlattice, and square.}
\label{fig3}
\end{figure}

We first focus on the sequence of bifurcations that occur with
increasing value of $f$ for a fixed value of ${\Delta}H/H_{c} = 0.22$.
The flat fluid surface becomes unstable first to a harmonic mode for a
small $f$.  Since the up-down symmetry is broken for a harmonic mode,
quadratic interaction is the most significant nonlinearity
\cite{ch93}, producing hexagonal standing wave pattern that is
synchronous to the ac driving [Fig.~\ref{fig1}(a) and
Fig.~\ref{fig3}(a)].  With a gradual increase in $f$, the wave number
($k_H$) increases slightly [see Fig.~\ref{fig3}(e)] but the hexagonal
symmetry does not change until $f=6.7$ Hz at which the system enters
into the bicritical region.  At the onset of bicriticality, 6
subharmonic peaks ($k_{\rm sub} = {k_{H} / {\sqrt{3}}}$) suddenly
appear forming a superlattice pattern together with the primary 6
harmonic peaks [see Fig.~\ref{fig1}(b) and Fig.~\ref{fig3}(b)].  
The transition from hexagon to superlattice is very sharp and appears 
to be first order, but the exact nature of the transition could not be
resolved within the accuracy of our current experimental setup.

\begin{figure}
\centerline{\psfig{figure=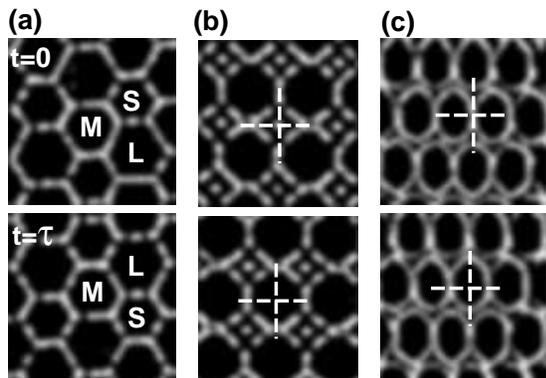,width=0.4\textwidth}}
\caption{Time evolution of subharmonic patterns during one driving
cycle: (a) superlattice, (b) square, and (c) rhombus.  The patterns
after one driving period ($\tau$) are shown in the bottom row.  The
dashed cross lines are placed to indicated the same location.}
\label{fig4}
\end{figure}

The observed superlattice pattern is similar but different from the
ones previously observed by others
\cite{Gollub98,Fineberg98,Knorr99-1,Knorr99-2}.  In particular, the
snapshot of superlattice [Fig.~\ref{fig4}(a)] taken at $t=0$ is
basically tiled by three different cells differing in sizes and
shapes.  During one driving cycle the cell ``M'' returns to the same
``M'', while ``S'' and ``L'' swap each other.  Upon completing the
subsequent driving cycle, ``M'' returns to the same, but ``S'' and
``L'' swap again.  In other words, the superlattice in
Fig.~\ref{fig1}(b) has a mixed mode oscillation.  This is in marked
contrast with the purely subharmonic hexagonal superlattices observed
by Wagner {\it et al.}~that alternates between two complementary
hexagonal lattices composed with cells of same type \cite{Knorr99-1},
and also with the ``Type-II superlattice'' observed by Kudrolli {\it
et al.}~that undergoes a complex stripe modulation during the course
of oscillation \cite{Gollub98}.

Since the superlattice forms in a narrow region between harmonic 
hexagon and subharmonic square, it is natural to believe that the
superlattice forms by an interplay between the two different modes.  
Since both harmonic and subharmonic modes are present, dominant
nonlinearity of inter-mode interaction is quadratic generating new
subharmonic modes.  In addition, translational invariance requires 
that the newly generated wavevectors are linear combinations of 
$k_{H}$ and $k_{SQ}$.  They would in turn interact with the dominant
harmonic hexagonal basis, and only those which satisfy the (spatial
and temporal) resonance conditions would survive \cite{Skeldon99}.
Two out of four subharmonic square modes $k_{SQ}$ and four out of new
subharmonic modes satisfy the resonance conditions, and form a
hexagonal sublattice with magnitude $k_{\rm sub} = k_{H}/ \sqrt{3}$,
thus contributing to the superlattice.  The present mechanism for the
generation of the superlattice is very similar to that of
\cite{Knorr99-2}.  

As $f$ is increased further, the superlattice undergoes a hysteretic
transition to a subharmonic square lattice; upon increasing
(decreasing) sequence of $f$, the superlattice (square) transforms to
the square (superlattice) at $f = 7.10$ Hz ($f = 7.20$ Hz).  At the
onset of transition to square lattice, $k_{SQ}$ is approximately
$k_{H}/\sqrt{2}$.  In the bistable regime, mixed states of square
domains and superlattice domains are often seen.  The observed square
patterns are subharmonic just like the ones that are popularly seen in
conventional Faraday experiments: the pattern alternates between two
dual square lattices at every driving period [see Fig.~\ref{fig4}(b)].
The $k_{SQ}$ increases significantly as $f$ increases.

Upon increasing $f$ further, the square lattice undergoes a hysteretic
transition to a stable rhombic pattern with a well defined 2-fold
symmetry [see Fig.~\ref{fig3}(d) and (e)].  At the transition, the
$k_{SQ}$ branch continues to $k_{RH1}$ (primary rhombic basis vector)
and a new basis vector $k_{RH2}$ that is the difference between two
neighboring $k_{RH1}$s is spontaneously created.  The oblique angle
between $k_{RH1}$ and $k_{RH2}$ characterizes a particular rhombic
pattern.  For the example shown in Fig.~\ref{fig3}(d),
$k_{RH2}/k_{RH1}=0.696$ and the angle between the two neighboring
basis wavevectors is $69.7^{0}$.  The oblique angle
changes but slightly as the driving frequency changes (less than $2\%$
for the frequency range 17.5 - 21.0 Hz).  The system is bistable for
the range ($17.5$ Hz $<$ $f$ $<$ $18.0$ Hz).  Except for the bistable
region, the rhombic lattice is globally stable.  Figure~\ref{fig4}(c)
clearly demonstrate that the observed rhombic pattern is temporally
subharmonic just like the square pattern shown in Fig.~\ref{fig4}(b).
As far as we know, the observed rhombic pattern is the first
laboratory demonstration of its kind ever revealed \cite{Swinney93}.

At this point, we note that a recent theoretical study has shown that
parametrically driven surface waves on ferrofluids can be unstable to
multiple spatial modes simultaneously \cite{Riecke97}.  Here, the
existence of multiple unstable modes originates from the non-monotonic
dispersion relation of ferrofluids.  The study has shown that in
one-dimensional geometry one can realize a situation that different
domains with different wavevectors coexist.  Such phenomenon is also
verified recently in an experiment conducted in a quasi
one-dimensional annular geometry \cite{Rehberg98-1}.  The existence of
rhombic pattern in our system, however, suggests that in
two-dimensional space the (two) unstable modes rather form a deformed
crystalline structure (i.~e. rhombic lattice), instead of domains with
different wavevectors.  In our experiments, coexisting domains with
different wavevectors are only observed in bistable regions.  In order
to check the feasibility of this hypothesis, we have extended the
earlier theoretical study by Raitt and Riecke \cite{Riecke97} to
two-dimensional space and have found a parameter regime in which well
defined rhombic patterns are globally stable \cite{model}.  Recently, Lifshitz
and Petrich have also demonstrated the existence of a stable rhombic
pattern in a generic model system containing two unstable spatial 
modes \cite{Petrich97}.

\begin{figure}
\centerline{\psfig{figure=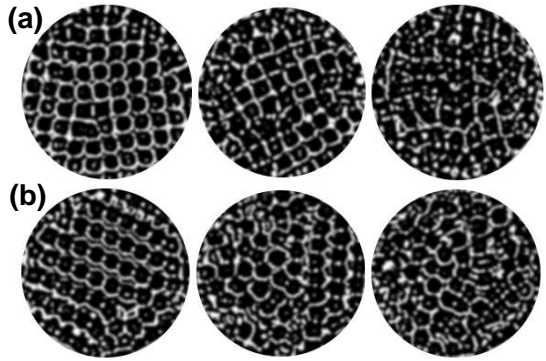,width=0.4\textwidth}}
\caption{Transitions to disordered patterns from (a) square and (b)
rhombus, as $\Delta H$ increases.  The frequency $f$ is fixed at
$15.0$ Hz for (a) and $21.0$ Hz for (b).  $\Delta H/H_{c}$ is $0.28,
0.39, 0.45$ (from left to right) for (a), and $0.28, 0.37, 0.45$ for
(b).}
\label{fig5}
\end{figure}

In the following, the transition to spatio-temporal chaotic state
(STC, see Fig~\ref{fig2}) from crystalline states is qualitatively
described.  As $\Delta H$ increases for a fixed value of $f$, the
crystalline patterns gradually become disordered.  Figure~\ref{fig5}
shows two such sequences, one for the case of square and the other for
the case of rhombic pattern.  Dislocations and defects start to form
in a square lattice near $\Delta H / H_{c}= 0.28$ [Fig.~\ref{fig5}(a),
left image].  As $\Delta H$ increases, the number of defects increases
(middle image) and eventually the whole lattice becomes very loose as
in a typical melting process.  The last image in Fig.~\ref{fig5}(a)
shows a state very near to STC, showing a very little trace of square
lattice.  A similar melting process is also observed for the case of
rhombic lattice [Fig.~\ref{fig5}(b)].  The gradual deformation toward
the state of STC is not limited to the square and rhombus but occurs
in the whole range of $f$ that we have explored.  This observation is
consistent with the earlier Faraday experiments conducted with
Newtonian fluids \cite{trg89}.

In conclusion, superlattice and rhombus, as well as square and
hexagonal standing waves, are observed on the surface of ferrofluids
driven parametrically with an ac magnetic field.  It is argued that
the superlattice pattern is formed by the resonance of newly generated
subharmonic modes with the harmonic hexagonal mode already present.
Due to the non-monotonic dispersion relation of the ferrofluids,
multiple spatial modes can become unstable simultaneously with a
single frequency forcing.  As a consequence, stable rhombic patterns
are possible and observed in the experiment.  Full understanding of
the observed patterns and their mutual boundaries, however, pose quite
a challenge.

We are indebted to W.~G.~Choe for many helpful discussions.  This work
is supported by Creative Research Initiatives of the Korean Ministry
of Science and Technology.

\end{document}